# Co-deformation Between the Metallic Matrix and Intermetallic Phases in a Creep-Resistant Mg-3.68Al-3.8Ca Alloy

M. Zubair[1,2], S. Sandlöbes-Haut[1], M. Lipińska-Chwałek[3,4], M. A. Wollenweber[1], C. Zehnder[1], J. Mayer[3,4], J.S.K-L. Gibson[1], S. Korte-Kerzel[1]

[1]Institute for Physical Metallurgy and Materials Physics, Kopernikusstr. 14, RWTH Aachen University, 52074 Aachen, Germany.

[2]Department of Metallurgical and Materials Engineering, G.T Road, UET Lahore, Pakistan.

[3]Central Facility for Electron Microscopy, RWTH Aachen University, Ahornstraße 55, 52074 Aachen, Germany.

[4]Ernst Ruska-Centre for Microscopy and Spectroscopy with Electrons (ER-C), Leo-Brandt-Str. 1 52428 Forschungszentrum Jülich, Germany.

## Abstract

The microstructure of Mg-Al-Ca alloys consists of a hard intra- and intergranular eutectic Laves phase network embedded in a soft α-Mg matrix. For such heterogeneous microstructures, the mechanical response and co-deformation of both phases under external load are not yet fully understood. We therefore used nano- and microindentation in combination with electron microscopy to study the deformation behaviour of an Mg-3.68Al-3.8Ca alloy.

We found that the hardness of the $Mg_2Ca$ phase was significantly larger than the α-Mg phase and stays constant within the measured temperature range. The strain rate sensitivity of the softer α-Mg phase and of the interfaces increased while activation volume decreased with temperature. The creep deformation of the $Mg_2Ca$ Laves phase was significantly lower than the α-Mg phase at 170 °C. Moreover, the deformation zone around and below microindents depends on the matrix orientation and is influenced by the presence of Laves phases. Most importantly, slip transfer from the α-Mg phase to the (Mg,Al)$_2$Ca Laves phase occurred, carried by the basal planes. Based on the observed orientation relationship and active slip systems, a slip transfer mechanism from the soft α-Mg phase to the hard Laves phase is proposed. Further, we present implications for future alloy design strategies.

# 1. Introduction

Increased fuel economy and the reduction of harmful emissions are the key factors behind a growing interest in the use of magnesium alloys for automotive applications [1-3]. Mg-Al-Ca alloys are among the most important alloys proposed for elevated- to high-temperature automotive applications (125-200 °C) such as powertrains [4]. They are well-known for their low weight, good creep resistance [5-8], and thermal stability [9] while maintaining good specific strength [4, 10].

Mg-Al-Ca alloys consist of a dual-phase microstructure: a soft α-Mg matrix reinforced with hard intermetallic Laves phases in the form of an interconnected network. Laves phases are a prominent class of intermetallic compounds with an $AB_2$ stoichiometry and a fairly complex crystal structure [11, 12]. Three types of Laves phases are known in Mg-Al-Ca alloys: $Al_2Ca$ (C15), $(Mg,Al)_2Ca$ (C36) and $Mg_2Ca$ (C14), which form during solidification as a result of the eutectic reactions: L → α-Mg + C36 and L → α-Mg + C14 (~514 °C, transformation temperature) [13]. The Ca/Al ratio controls the type(s) of Laves phase(s) formed in the microstructure [14-16]. In alloys with a high Ca/Al ratio (~1), the Laves phases formed in the as-cast microstructure are either $(Mg,Al)_2Ca+Mg_2Ca$ (as in the present work) or $Mg_2Ca$ (if the ratio is significantly higher than ~1). At room temperature (RT), the critical resolved shear stress (CRSS) for basal slip in Mg amounts to < 1 MPa [17], while the CRSS are < 10 MPa for tensile twinning [18, 19], ≈ 39 MPa for prismatic slip [20] and ≈ 40-44 MPa for pyramidal slip [21, 22]. These values change with temperature and composition (addition of alloying elements) [19, 20, 23-27]. The CRSS values of basal, prismatic and pyramidal slip in Laves phases are well above the values of pure and alloyed Mg; e.g. for the $Mg_2Ca$ phase, the CRSS for basal slip has been reported based on micro-compression experiments to amount to ≈0.52 GPa, for prismatic slip ≈0.44 GPa and ≈0.53 GPa for $1^{st}$ order pyramidal slip [28]. The hardness (related to the resistance to dislocation glide) of the other Laves phases in the Mg-Al-Ca system is even higher than that of the Mg$_2$Ca Laves phase [29]. The difference, therefore, between the CRSS required to activate basal, prismatic, or pyramidal slip in both phases amounts to at least two orders of magnitude (see Fig. SM 1). Under external loading, it is then expected that the soft α-Mg matrix will deform first. The Laves phase skeleton is also expected to restrict the dislocation movement in the matrix and will correspondingly affect the deformation occurring in the soft phase.

In order to improve the alloys' deformation and failure behaviour in application, it is essential to understand the co-deformation behaviour of the two, mechanically very different, phases (soft α-Mg and hard Laves phases) under external loading at different temperatures and time scales.

Therefore, the present study addresses the following key aspects related to the deformation behaviour of a dual phase Mg-3.68Al-3.8Ca alloy:

1. Mechanical properties of the individual phases and thermally activated deformation mechanisms, determined over a temperature range spanning from room temperature to an application temperature of 170 °C,
2. The effects of matrix orientation and presence of Laves phases on the deformation of the α-Mg matrix,
3. Slip transfer from the soft α-Mg phase to the hard intermetallic Laves phase.

All these aspects were studied using a combination of nano and microindentation, scanning electron microscopy (SEM), electron backscatter diffraction (EBSD) and transmission electron microscopy (TEM).

## 2. Experimental methods

A permanent mould cast Mg-3.68Al-3.8Ca (referred to as AX44 throughout the paper) and a master alloy Mg-30Ca consisting of large areas of $Mg_2Ca$ Laves phase (both in wt.%, here and below) were investigated in this study. The melting and casting procedures of alloy AX44 are described in [30], the master alloy was investigated in its as-received state. Separate disc-shaped specimens (~12.5 mm diameter, ~1 mm thickness) for nano- and microindentation testing were machined from the as-cast block and the master alloy by electric discharge machining. The specimen preparation routine used to prepare samples for microscopy, nano- and microindentation are described in [30, 31].

The micro-/nano-mechanical behaviour of the alloy AX44 was investigated by nanoindentation (Nanomechanics InSEM-III) using a diamond Berkovich indenter tip (Synton MDP, Switzerland) and a continuous stiffness measurement (CSM) unit. The diamond area function (DAF) and frame stiffness were calibrated on a standard fused $SiO_2$ sample at room temperature according to the Oliver and Pharr method [32]. The indenter tip and specimen were heated separately to minimise thermal drift, which was stabilised below 0.1 nm/s. Nanoindentation testing was carried out within the temperature range from RT to 170 °C. Three different types of nanoindentation tests were performed on alloy AX44: i) constant strain rate (CSR) tests, ii) strain rate jump (SRJ) tests and iii) nanoindentation creep (relaxation) tests. All the nanoindentation tests on the AX44 alloy were performed on the same sample in a way that the sample was first heated up to a temperature of 170 °C using an in-situ nanoindenter and then all three types of tests were performed. Afterwards the temperature was decreased to 140 °C, and the same tests were done on a different location. Further tests at 100 °C and room temperature were carried out subsequently. As incorporation of Al, as evidenced by a reduction in lattice parameters and the presence of C15 stacking faults was reported for the C36 Laves phase at high temperature and after 100 or more hours [33, 34], we performed the experiments all on the same sample with descending temperature to minimize the effect of this change in the C36 Laves phase on the deformation behaviour at different temperatures. Due to the increased size of the indents in the nanoindentation creep tests, these were also performed separately on the Mg$_2$Ca Laves phase present in the Mg-30Ca master alloy (Fig. SM 2). Besides that, CSR tests were also performed on the Mg-30Ca master alloy to determine the hardness of the Mg$_2$Ca phase over a range of temperature. High temperature CSR tests on Mg$_2$Ca phase were also conducted in the similar manner like mentioned for the AX44 alloy. The hardness obtained for the Mg$_2$Ca phase is an average of three different orientations. As these were the only tests performed on this sample, unless otherwise stated, the results presented here refer to the AX44 material.

CSR tests were carried out at a strain rate of 0.2 s$^{-1}$ to a maximum load of 45 mN. The load was held constant at maximum load for one second before unloading. CSR, SRJ, and nanoindentation creep tests were performed at RT, 100 °C, 140 °C and 170 °C. During the SRJ tests, the strain rate was varied between 0.1 s$^{-1}$ and 0.001 s$^{-1}$ in two jumps between indentation depths of 500-800 nm. The creep tests were performed at a load of 16 mN, applied at a rate of 160 mN/s, which was then held constant for 120 seconds before unloading.

Microhardness tests were carried out using a Vickers hardness tester at a force of 0.5, 5 and 10 N at room temperature (RT) on alloy AX44. The holding time at maximum load for all micro indents was 15 seconds. The deformed regions below the indents were studied using the bonding-interface technique [35-37]. Flat, rectangular specimens (~7 x ~4 x ~3 mm, machined and metallographically prepared in the same way as mentioned for disc specimens) were used for this purpose. The polished surfaces of two flat specimens were glued together (using Crystal bond) and then the indents were made at the interface. The samples were then placed in acetone in order to dissolve the glue.

Microstructure characterisation was performed using secondary electron (SE) and back-scattered electron (BSE) imaging at 10-20 kV in a scanning electron microscope (SEM) (FEI Helios Nanolab 600i and Zeiss LEO1530). Electron backscatter diffraction (EBSD, FEI Helios Nanolab 600i) was carried out at an acceleration voltage of 20 kV.

Site specific preparation of the electron transparent samples for transmission electron microscopy (TEM) was performed with the aid of focused ion beam milling (FEI Helios Nanolab 600i). The FIB lamellas were prepared in such a way that the viewing direction was perpendicular to the indentation direction (the top of the lamellas is shown outlined in Figure 6 and Figure 9). TEM observations were performed at 200 kV (FEI Tecnai G2 F20)[38].

## 3. Results

### 3.1. Microstructure of the as-cast alloy

The as-cast microstructure of alloy AX44 is presented in Figure 1 (a) and (b) with the aid of SEM-BSE images. The interconnected network of intermetallic Laves phases has been studied using SEM, TEM and energy dispersive X-ray spectroscopy (EDS) by other researchers [13, 39, 40] and found to consist of a mixture of two different morphologies in similar alloy systems (having comparable Ca/Al ratios). These correspond to two different Laves phases: i) coarse eutectic $(Mg,Al)_2Ca$ and ii) fine eutectic $Mg_2Ca$. In addition to intermetallic network, thin needle like $Al_2Ca$ precipitates were also observed in the as-cast alloy in line with the previous work [41-43]. The EDS analysis and further details about the microstructural constituents of alloy AX44 are presented in a previous paper [30].

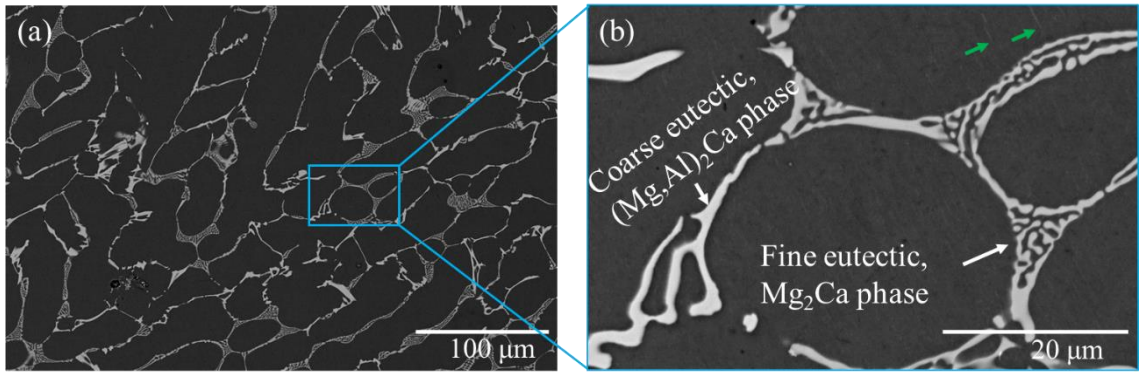

*Figure 1: Microstructure (SEM-BSE) of as-cast AX44 showing an interconnected Laves phase network consisting of coarse eutectic (Mg,Al)$_2$Ca and fine eutectic Mg$_2$Ca. White arrows depict the intermetallic Laves phase while green arrows show the presence of Al$_2$Ca precipitates within the α-Mg matrix.*

### 3.2. Nanoindentation

#### 3.2.1. CSR tests

**3.2.1.1. Hardness variation of all three microstructural constituents with temperature**

Figure 2 (a) shows the average hardness (at a depth of ~500 nm) of the α-Mg and Mg$_2$Ca phases from RT up to 170 °C and 200 °C, respectively. The scatter from averaging the values of the α-Mg phase at all temperatures is so small that it is hardly visible in Figure 2 (a) (also given in Table 1). The hardnesses of a few randomly-selected indents at the interface are also represented in Figure 2 (a) as red triangles. As the volume fractions vary for indents placed at the interface (see Fig. SM 3), individual data points are shown rather than an average and its deviation. The hardness of α-Mg phase decreases slightly while that of Mg$_2$Ca phase stays constant with temperature. The hardness at the interfaces is significantly higher than the hardness of the α-Mg matrix at all test temperatures. A small indentation size effect is apparent in the deformation of the α-Mg matrix at all temperatures, i.e, the hardness of the α-Mg matrix decreases with increasing depth (till ~1000nm) (see Figure 2 (b)). The indentation size effect (i.e. a decreasing hardness with increasing depth) is a well-known phenomenon in the indentation of crystalline materials and is attributed to the strain gradients produced beneath indents and the related density of geometrically necessary dislocations [44-46]. For the indents in the magnesium matrix, there is a slight increase in the mean hardness after a depth of ~1000nm was reached because of the hardening effect from interfaces (Figure 2 (b)).

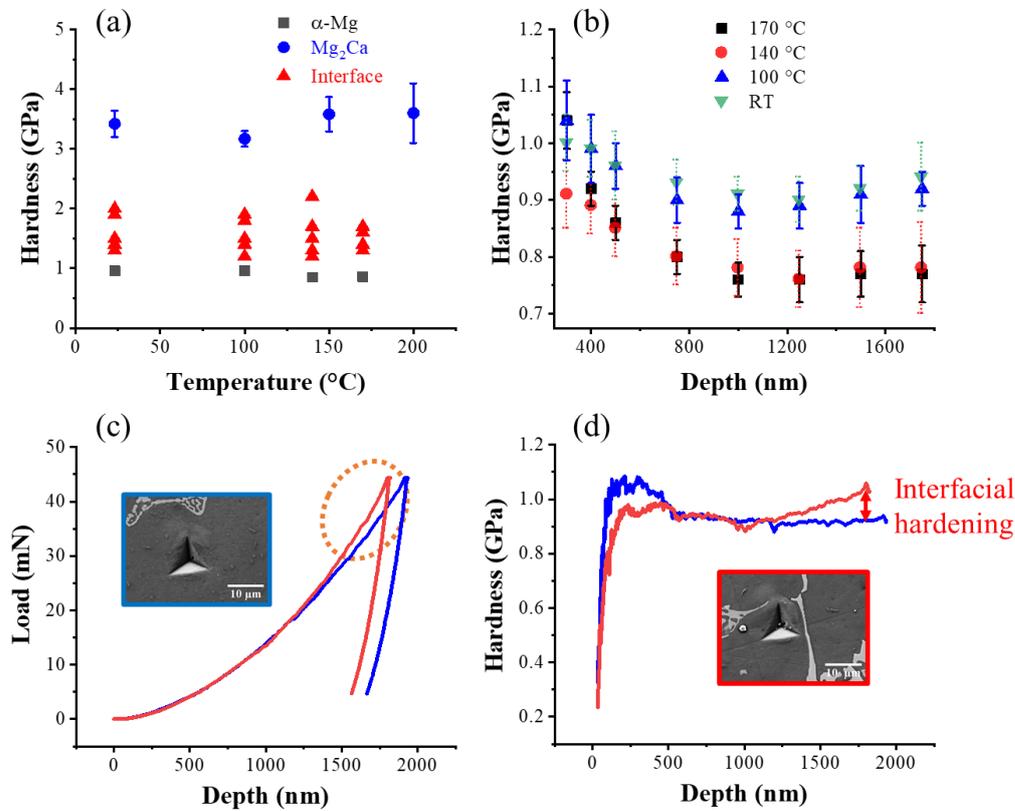

*Figure 2: (a) Average hardness of the α-Mg phase, Mg₂Ca phase and at the interface up to a temperature of 170 °C (for Mg₂Ca till 200 °C), (b) dependence of the hardness on the indentation depth for α-Mg matrix at all test temperatures, (c) Load-depth, and (d) hardness-depth curves (at RT) highlighting the effect of interfaces on the depth dependent hardness. Hardness of the Mg₂Ca phase at RT was taken from [28].*

### 3.2.1.2. Hardening from interfaces

Load-depth and hardness-depth curves of two different indents made at RT are given in Figure 2 (c and d). The blue curves are from an indent made some distance away from a Laves phase, the red curves from an indent close to one. The hardness of both indents is nearly the same between 500-1000 nm, beyond which there is an increase in hardness for the indent made closer to the α-Mg/Laves phase interface (shown red). This increase in hardness is assumed to be because the interface blocks dislocation motion.

### 3.2.2. SRJ Tests

Nanoindentation SRJ tests were performed to determine the strain rate sensitivity (SRS) of individual phases and other microstructural features, such as interfacial areas, in order to gain a deeper understanding of the thermally activated deformation mechanisms taking place in these regions. The strain rate sensitivity, *m*, was calculated according to [47-49]

$$m = \frac{\partial lnH}{\partial ln\dot{\varepsilon}} \qquad \text{Eq. 1}$$

where *H* is the hardness and $\dot{\varepsilon}$ is the indentation strain rate. $\dot{\varepsilon}$ can be determined using the true strain rate concept using [47, 50]:

$$\dot{\varepsilon} = \left(\frac{1}{h}\right)\left(\frac{dh}{dt}\right) = \frac{\dot{h}}{h} = \frac{1}{2}\left(\frac{\dot{P}}{P} - \frac{\dot{H}}{H}\right) \approx \frac{1}{2}\frac{\dot{P}}{P} \qquad \text{Eq. 2}$$

where *h* is the contact depth, and *P* is the applied load. $\dot{h}$, $\dot{P}$ and $\dot{H}$ are the respective rates (with respect to time) of depth, load, and hardness.

The activation volumes, *V\**, were calculated using [47, 48, 51-54]

$$V^* = 3\sqrt{3}kT\left(\frac{\partial ln\dot{\varepsilon}}{\partial H}\right) \qquad \text{Eq. 3}$$

where *k* is the Boltzman constant and *T* is the absolute temperature.

The hardness-depth and modulus-depth curves are shown in Figure 3 (a) and (b), for two representative indents performed at 170 °C within the α-Mg phase and at the α-Mg/Laves phase interface, respectively. Both microstructural regions showed an indentation size effect below 1000 nm, therefore, linear fitting was carried out for each strain rate jump in the hardness-depth curve. The instantaneous hardness values (measured right before (H($\dot{\varepsilon}_1$)) and after the strain rate jump (H($\dot{\varepsilon}_2$)) as depicted in the magnified region of Figure 3 (a)) were then used to calculate *m* and *V\** as suggested by Maier et al. [45] and Wu et al. [55].

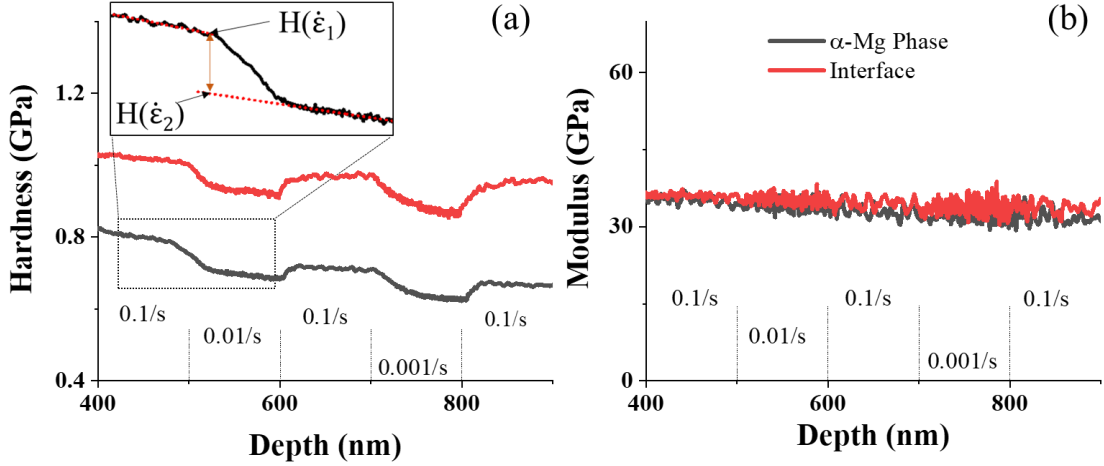

*Figure 3: a) Variation of hardness with strain rate of the α-Mg phase and α-Mg/Laves phase interfaces at 170 °C and b) modulus vs depth curves with strain rate for the α-Mg phase and α-Mg/Laves phase interfaces at the same temperature. $H(\dot{\varepsilon}_1)$ is the value of hardness before the strain rate jump and $H(\dot{\varepsilon}_2)$ represents the value of hardness directly after the strain rate jump, found by extrapolating the hardness-depth curve backwards once the transient had stabilised.*

The magnitude of the hardness change with strain rate in the α-Mg phase increased with temperature. Consequently, the value of *m* also increased and amounts to 0.004 at RT and 0.023 at 170 °C (details in Table 1). The indents made across α-Mg-Laves phase interfaces revealed that this microstructural region exhibits a similar strain rate sensitivity as that of the α-Mg phase at all testing temperatures, despite the significantly increased hardness. The values of the strain rate sensitivity were calculated from a minimum of eight indents at all testing temperatures (except RT, where six indents were used to calculate *m*). A size effect in *m* was observed for the α-Mg phase (see Fig. SM 4) and the *m* values given in Table 1 are average values at all strain rate jumps in between 500-900 nm.

*Table 1: Hardness at a depth of 500 nm, m and V* values of the α-Mg phase and α-Mg/Laves phase interfaces at different temperatures.*

| Temp (°C) | Hardness (α-Mg), [GPa] | Strain rate sensitivity, *m* | | Activation volume, $V^*$ [nm$^3$ (b$^3$)] | |
|---|---|---|---|---|---|
| | | α-Mg phase | Interface | α-Mg phase nm$^3$ [b$^3$] | Interface [nm$^3$] |
| RT | 0.96±0.06 | 0.004±0.002 | 0.006±0.001 | 7.66 ± 3.70 (232 ± 112) | 3.35 ± 0.91 |
| 100 | 0.96±0.04 | 0.013±0.006 | 0.016±0.007 | 2.84 ± 1.12 (86 ± 34) | 1.48 ± 0.82 |
| 140 | 0.85±0.05 | 0.016±0.006 | 0.020±0.006 | 2.63 ± 1.18 (79 ± 36) | 1.27 ± 0.40 |
| 170 | 0.86±0.03 | 0.023±0.009 | 0.024±0.008 | 2.31 ± 0.87 (70 ± 26) | 1.46 ± 0.54 |

Table 1 lists the values of the activation volume, $V^*$, for the α-Mg phase and α-Mg/Laves phase interfaces in nm$^3$ and b$^3$, respectively, when considering the <a> Burgers vector of Mg (3.21 x 10$^{-10}$ m) [54, 56] at various temperatures. There is a significant decrease in $V^*$ from RT to 100 °C, but with a further increase in temperature up to 170 °C it remains relatively constant (within the measurement noise) for both the α-Mg phase and α-Mg/Laves phase interfaces. In general, however, the α-Mg/Laves phase interfacial regions possessed lower activation volumes than the α-Mg phase at all testing temperatures. Several researchers [51, 57, 58] have used modulus compensated hardness to calculate activation energies, however, in this work the modulus decreased at the nearly same rate as the hardness which precludes the calculation of the activation energy and thermally activated mechanisms in this manner. We therefore rely solely on the activation volumes determined at each temperature.

### 3.2.3. Nanoindentation creep testing

As the investigated alloys are designed for elevated-temperature applications, nanoindentation creep tests were also performed on alloy AX44 and the Mg-30Ca master alloy (consisting of large areas of Mg$_2$Ca Laves phase) at room temperature, 100 °C, 140 °C and 170 °C. Figure 4 shows the fitted creep curves of the α-Mg phase (present in alloy AX44) at various temperatures. The creep rate in indentation creep testing can be related to the nominal pressure ($p_{nom}$ =P/A, where P is the applied load and A is the contact area of indentation) or hardness according to Eq. 4 [59, 60]

$$\dot{\varepsilon} = \left(\frac{1}{h}\right)\left(\frac{dh}{dt}\right) = B\,(p_{nom})^N \qquad Eq.\ 4$$

with *h* being the indentation depth, *B* a material constant, and *N* the stress exponent obtained from nanoindentation.

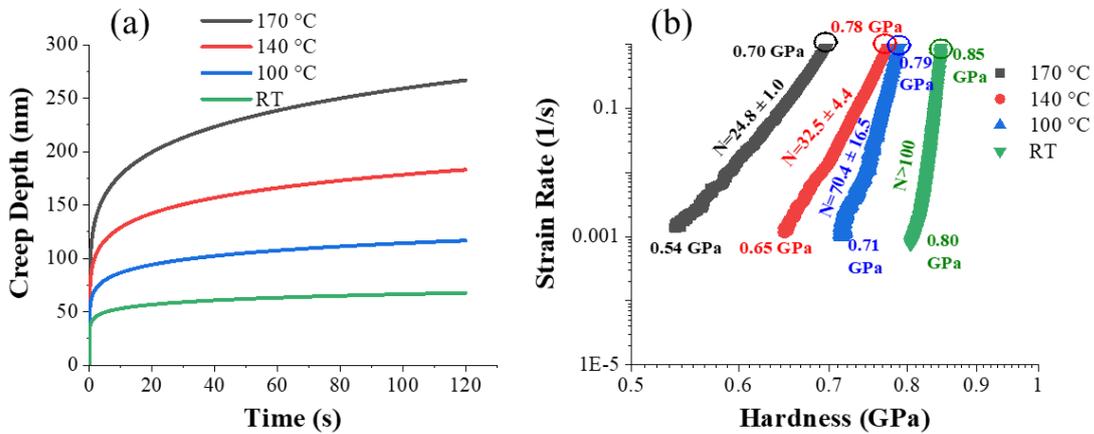

*Figure 4: (a) Creep depth as a function of temperature of the α-Mg phase (present in alloy AX44); (b) strain rate vs hardness of the α-Mg phase at various temperatures, the first data point in the graph is highlighted by open circles.*

The creep depth increased, and the stress exponent decreased with increasing temperature, Figure 4 (a) and (b), respectively. The nanoindentation stress exponent of this alloy was nearly four times higher than the stress exponent obtained previously from uniaxial tensile testing (n = 5.42) at the same temperature of 170 °C [30]. This discrepancy between (nano)indentation creep and macroscopic creep tests is relatively common [61, 62], and can largely be explained by the fact that the stress and strain levels in nanoindentation using a Berkovich indenter are significantly higher than that during macroscopic uniaxial tensile testing. Accordingly, it is generally assumed that the hardness resulting from a Berkovich indenter corresponds to a flow stress at 8% strain [57, 63]. A similar difference between the stress exponents obtained from nanoindentation creep tests and macroscopic creep tests was also observed by Wu et al. [61].

Nevertheless, the indentation creep tests allowed a comparison between the α-Mg (present in the AX44 alloy) and $Mg_2Ca$ (present in the Mg-30Ca master alloy) phases at the same load (16 mN) and temperature (170 °C) (Figure 5 (a and b)), showing that creep deformation was nearly 4.5 times lower in the $Mg_2Ca$ phase than in the α-Mg phase. Moreover, the hardness of the $Mg_2Ca$ phase is significantly higher than that of the α-Mg phase. There exists an elegant numerical agreement between creep depth and hardness of both phases. However, caution must be taken while using short term hardness tests as a proxy for more laborious indentation creep testing, at least for determining relative behaviour between phases. This is because sometimes the rate of hardness loss changes with increasing or decreasing strain rate as evident in Fig. SM 5 and as reported by Mathur et al. [57] while working with $Mg_{17}Al_{12}$ phase.

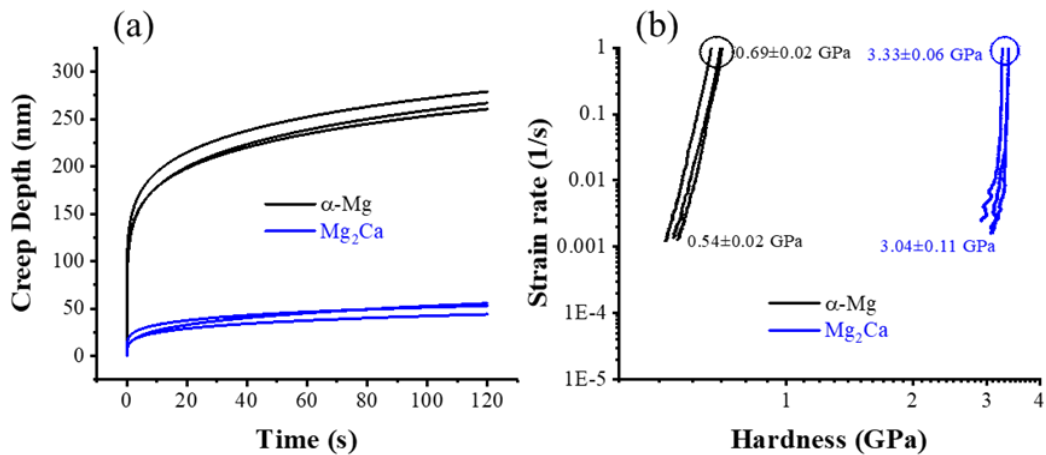

*Figure 5: (a) Nano indention creep curves for the α-Mg phase and the $Mg_2Ca$ phase (present in the Mg-30Ca master alloy), (b) Strain rate-hardness curve for α-Mg and the $Mg_2Ca$ phase, the first data points are highlighted with open circles.*

### 3.3. Microindentation

The microhardness of the as-cast AX44 at 5 N was determined as 61.2 ± 3.5 HV. In line with previous work [30], cracks in the Laves phase were observed within the direct proximity of the indents. In addition, parallel slip lines were observed on the sample surface around the indent (Figure 6). Coincident positions of the slip lines at the interface between the α-Mg matrix and the (Mg,Al)$_2$Ca phase (see Figure 6) indicate that co-deformation of both phases has occurred.

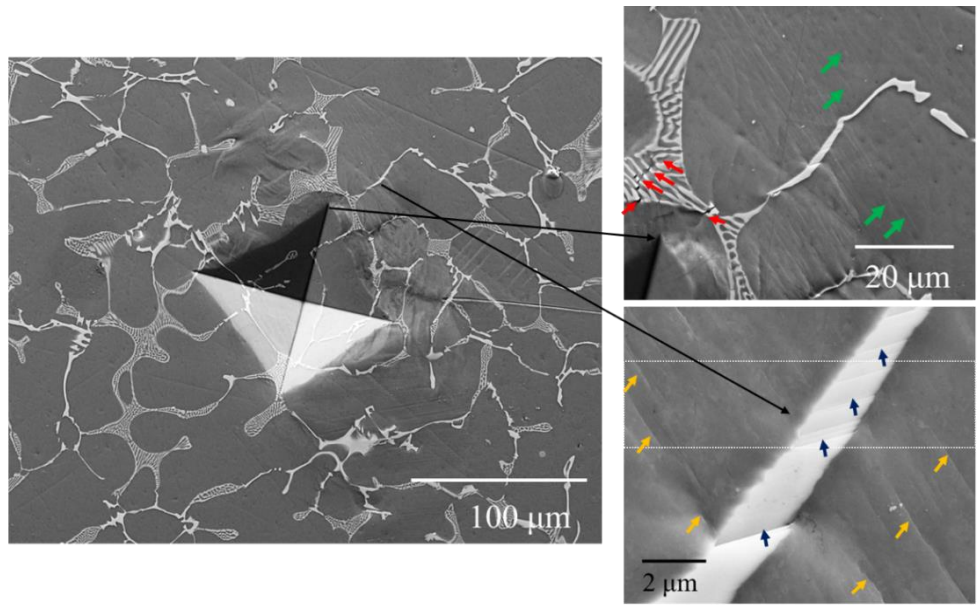

*Figure 6: SEM (SE) images of a microindent at 5N performed on alloy AX44. The magnified regions are shown as insets. The red arrows indicate cracks in the Mg$_2$Ca and (Mg,Al)$_2$Ca Laves phases, green arrows highlight Al$_2$Ca precipitates, orange arrows show slip lines in the α-Mg phase and blue arrows depict slip lines in the (Mg,Al)$_2$Ca Laves phase. The dotted white rectangle indicates the region of the FIB lamella cut.*

As is expected from a hexagonal metal, the deformation zone in the α-Mg phase is strongly orientation-dependent, i.e. the presence of twins and slip lines around indents are significantly affected by the orientation of the α-Mg phase. Figure 7 shows an inverse pole figure (IPF) map superimposed with band contrast (BC) map (also known as pattern quality) together with the corresponding SE micrographs of microindents in alloy AX44. It can be seen in Figure 7 (a-d) that the appearance of the area around the microindents is orientation dependent. Further, formation of cracks in the Laves phase was observed at places where slip lines in the α-Mg phase intersect with the Laves phase, as indicated in Figure 7 (e) and (f), which is in line with observations reported in previous work [31], where cracks were observed at similar regions in macroscopic uniaxial tensile deformation.

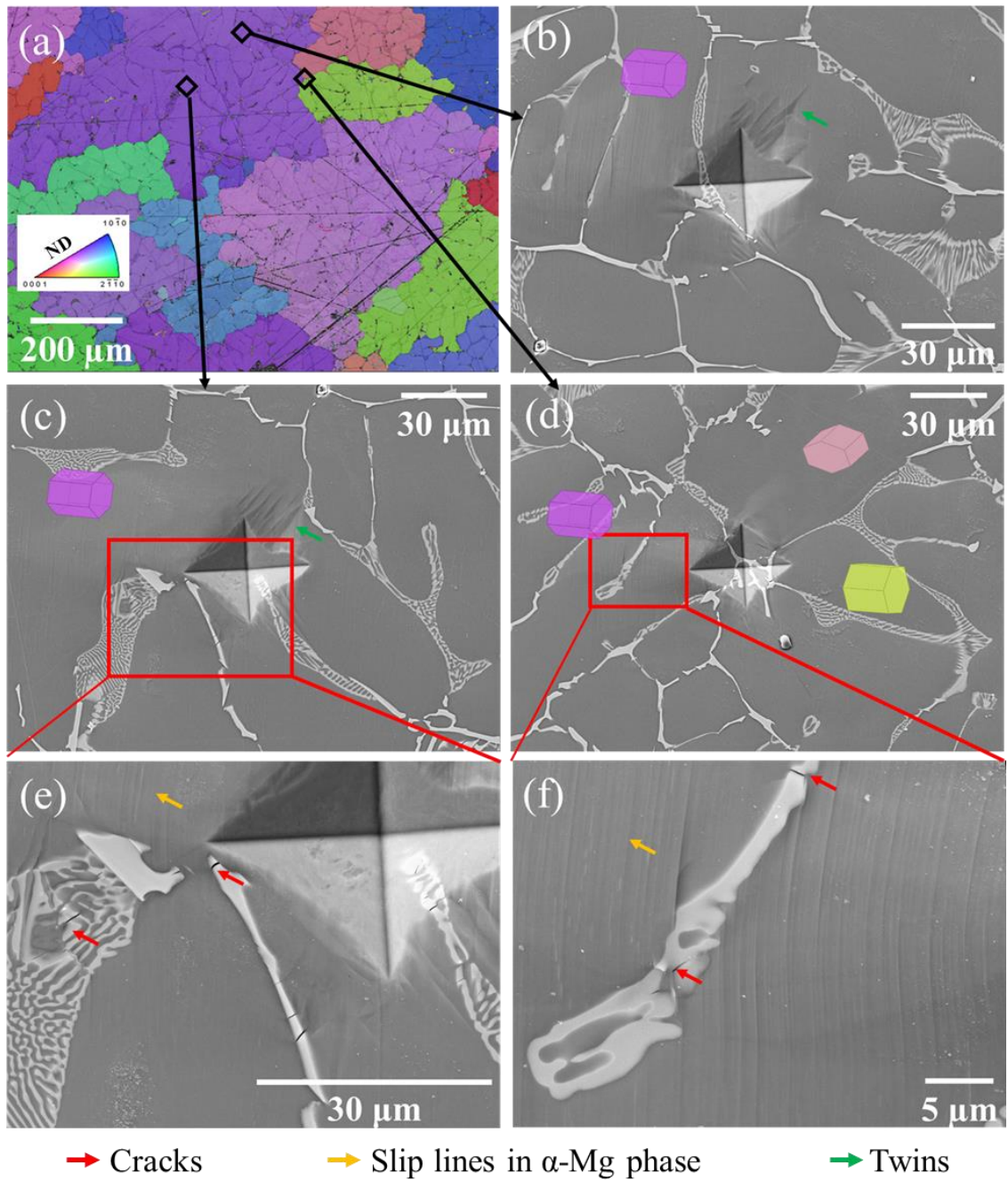

Figure 7: Microstructure and deformation behaviour in alloy AX44. (a) Inverse pole figure (IPF) and band contrast (BC) maps superimposed on each other (black arrows indicate positions of the indents presented in detail in (b)-(d)), (b) and (c) SE images of 500 mN indents in one grain with twin formation around the indentations, (d) SE image of an indent made at a grain triple point. (e) and (f) show magnified images of the regions indicated in (c) and (d), respectively.

The presence of Laves phases further modifies the shape of the deformation zone. Figure 8 (a) shows superimposed IPF and BC maps of the alloy AX44, with microindents placed at two different locations within the same grain: near the $(Mg,Al)_2Ca$ Laves phase (Figure 8 (b)) and away from the Laves phase network (Figure 8 (c)). The indentation-induced deformation twinning of the α-Mg matrix, (as confirmed by post-deformation EBSD) was observed around the indent (indicated with a dashed ellipse in Figure 8 c),) in the Laves-phase-free area, but deformation twinning was suppressed in the vicinity of the hard Laves phase despite the same crystal orientation and the same relative position to the indent (see Figure 8 (b)). This demonstrates the importance of co-deformation studies, as there is a clear effect of the intermetallic skeleton.

Finally, it was observed that in addition to the previously shown co-deformation resulting from dislocation slip in the α-Mg phase (Figure 6), similar co-deformation can be introduced from other deformation structures, as indicated by the green arrow in Figure 8 (d). Here, the deformation mechanism in the α-Mg is most likely deformation twinning (only one small twin like region is present, assumed from visual analysis as it was not possible to confirm this as a twin by EBSD, as compared to multiple twins present in Figure 8 (c)), but it is nevertheless present adjacent to slip lines in the Laves phase.

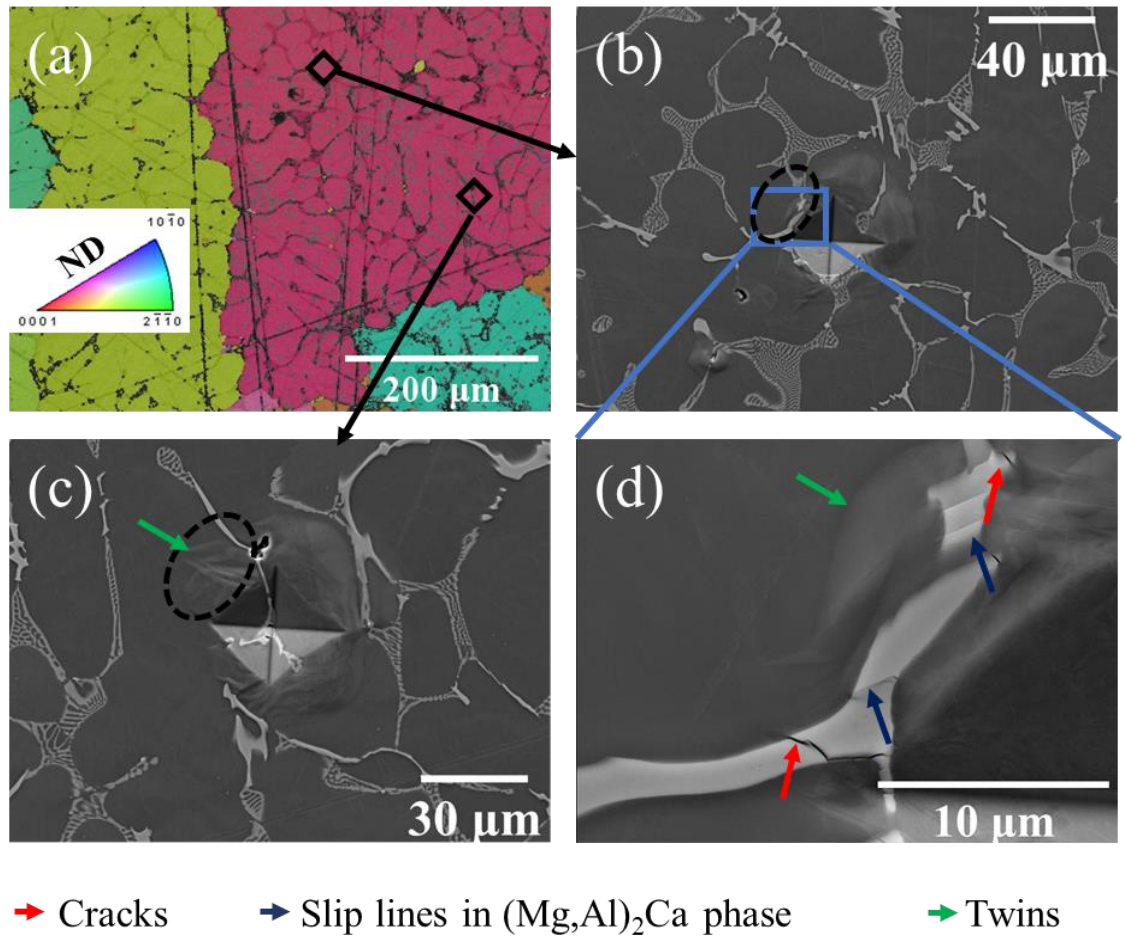

*Figure 8: (a) Superimposed IPF and BC maps of a grain containing two microindents. The position of the indents in the grain is highlighted by black squares and arrows in (a); (b) and (c) show the SE images of the two indents (made at 500 mN). The black dotted ellipses highlight differences in deformation behaviour. (d) Shows the magnified region highlighted by the box in (b). Twins in the α-Mg phase are represented by green arrows, red arrows depict cracks and blue arrows show slip lines in the (Mg,Al)$_2$Ca Laves phase.*

Having demonstrated the effect of crystal orientation and Laves phases on the deformation microstructure visible around the indents at the indented sample surface (Figure 7 and Figure 8), the sub-surface deformation is now presented. These studies are intended to determine whether the deformation observed in the presence of a free surface are representative for the mechanisms that occur in the bulk.

Figure 9 (a-d) shows SE images of the deformed sub-surface regions of the indents, produced at 5 (a,b) and 10 N (c,d)). Figure 9 (a) and (b) highlight dislocation slip (orange arrows) in the α-Mg phase as well as cracking in the Laves phase (red arrows). The curved slip lines, as indicated in the magnified inset in micrograph (c), are indicative of multiple active slip systems in the α-Mg phase. Figure 9 (d) demonstrates co-deformation of the α-Mg and the Laves phases, evident from the parallel slip lines in the Laves phase (blue arrows) coincident with the slip lines in the matrix phase (orange arrows). Overall, the observed surface deformation around the indents (Figure 6, Figure 7 and Figure 8), including the mix of co-deformation and cracking in the Laves phase, do indeed appear to be representative for deformation of the bulk material.

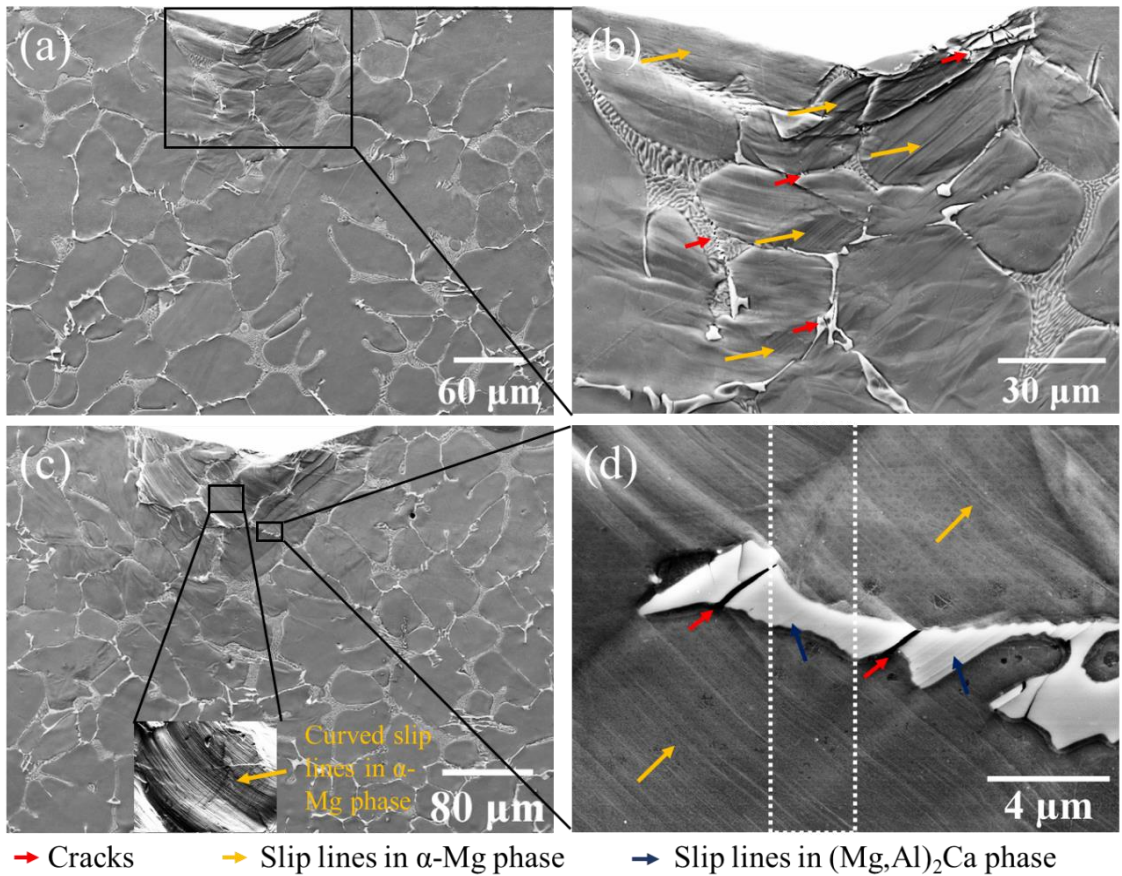

*Figure 9: SEM (SE) micrographs of the sub-surface region of indents at 5 N (a-b) and 10 N (c-d). Overview of the sub-surface features from an indent made at 5 N (a). (b) Represents the magnified image of the area highlighted with a black rectangle in (a). Overview of similar sub-surface features from an indent made at 10N (c). Inset in (c) highlights the curved slip lines observed in the α-Mg matrix. (d) Reveals co-deformation of the α-Mg matrix and the Laves phases. Orange and blue arrows indicate the slip lines in α-Mg and Laves phases, respectively, while cracks are highlighted by red arrows. The white dotted rectangle shows the region from where the TEM lamella was lifted out.*

Cracks and slip transmission events from the α-Mg phase into the C36 Laves phase were observed under all three loading conditions (Figure 6, Figure 8 and Figure 9). However, higher and more frequent instances of slip transmission were detected under higher loading, i.e., 10 N (see Figure 9 and Fig. SM 6). In the sub surface analysis presented in Fig. SM 6, more instances of slip transfer are evident for the indent made at 10N when compared to the indent made at 5N. This is because the higher force in indentation generally results in high plastic strain across a larger volume, the plastic zone, causing the aggregation of high-densities of dislocations at the Mg/Laves phase interface. Therefore, higher loads seem to promote more slip transmission from the α-Mg to the C36 Laves phase.

### 3.4. TEM

Figure 10 shows bright field (BF) TEM and corresponding selected area electron diffraction (SAED) images of the microstructure observed in a FIB lamella cut from the surface as indicated in Figure 6 by a white dotted rectangle (located aside the indent - FIB cut plane parallel to the indentation direction and perpendicular to the sample surface). The slightly curved volume of C36 Laves phase in the lamella is surrounded by a single Mg matrix (α-Mg) grain.

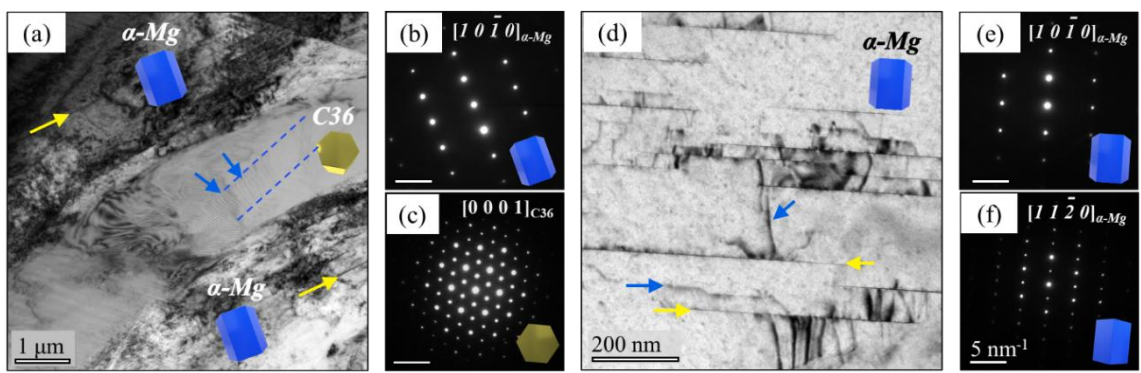

*Figure 10: Composite microstructure of the lamella as observed in TEM: a) BF overview image acquired with lamella orientation B ∥ $[1\ 0\ \bar{1}\ 0]_{α\text{-}Mg}$ with corresponding SAED of Mg matrix given in (b). c) SAED of the C36 grain shown in (a), observed in orientation B ∥ $[0001]_{C36}$. d) Magnified BF appearance of the Mg matrix structure with edge-on basal-plane SFs (yellow arrows) and non-basal dislocations (blue arrows). e) - f) SAED images of the Mg matrix shown in (d), acquired with sample orientation B ∥ $[1\ 0\ \bar{1}\ 0]_{α\text{-}Mg}$ and B ∥ $[1\ 1\ \bar{2}\ 0]_{α\text{-}Mg}$. Matrix and C36 orientation are visualised by blue and yellow hexagons, respectively. Images in (d-f) contain a 45° clockwise relative sample rotation with respect to those in (a-c).*

The crystallographic orientation of both phases was determined using SAED. The Mg matrix located on either side of the C36 belongs to the same matrix grain exhibiting, accordingly, almost the same orientation (see Figure 10 (a)), with negligible variation (below 2° across the lamella) related mainly to the bending of thin samples. The C36 phase grain also exhibits the same crystallographic orientation across the FIB lamella (less than 2.5° misorientation). Analysis of the relative crystallographic orientation of both phases based on corresponding SAED data (see Figure 10 (b) and (c)), revealed an almost perpendicular orientation (within ± 3°) of their basal planes. In particular, ∠[1 $\bar{2}$ 1 0]$_{C36}$, [0 0 0 1]$_{α-Mg}$=3° and [1 0 $\bar{1}$ 0]$_{C36}$ ∥ [3 $\bar{4}$ 10]$_{α-Mg}$. BF images in Figure 10 (a) and (d), show the microstructure observed with the electron beam direction B ∥ [1 0 $\bar{1}$ 0]$_{α-Mg}$. Stacking faults (SFs) in the α-Mg phase are observed in (a) and (d) along (or very close to) [1 0 $\bar{1}$ 0]$_{α-Mg}$. The crystallographic direction in an edge-on orientation (with the SF habit plane perpendicular to the image plane and parallel to the viewing/beam direction), reveals dark, thin, and straight lines (indicated by yellow arrows). Identification of SFs in the overview (a) is rather difficult due to strong diffraction contrast present in the Mg grain (dark, blurry appearance), resulting from the in-axis grain orientation (i.e. observation condition with exact B ∥ [1 0 $\bar{1}$ 0]$_{α-Mg}$). Accordingly, a better presentation of the edge-on SFs in the Mg matrix is shown in the magnified BF image presented in (d), where diffraction contrast is lowered by slight sample tilt (~ 2° away from the diffraction condition B ∥ [1 0 $\bar{1}$ 0]$_{α-Mg}$ ). The corresponding SAED pattern for the Mg grain in the orientation presented in (a) is shown in (b), while the SAED pattern for the [1 0 $\bar{1}$ 0]$_{α-Mg}$ condition, i.e. close to the observation conditions applied in (d) is shown in (e). Figure 10 (f) presents an SAED pattern of the other main zone axis of the hexagonal Mg [1 1 $\bar{2}$ 0]$_{α-Mg}$, acquired with the sample tilted 30° away from the orientation conditions applied in (d).

The presence of dislocations on a non-basal plane was also observed in the α-Mg matrix (e.g. curved dark lines, indicated in Figure 10 (d) with blue arrows), evidencing the activation of a non-basal deformation system within the Mg matrix of the composite material. Within the C36 grain, dislocation contrast confined within defined slip planes was also observed (indicated by blue arrows in Figure 10 (a)). Due to an unfavorable orientation of the slip planes almost parallel to the lamella plane, an edge-on observation and unambiguous confirmation of the crystallographic orientation of the habit plane was not possible in this sample. Nonetheless, a basal-slip co-deformation was postulated for the composite Laves phase taking into account the crystallographic orientation of the C36 grain, with $[0001]_{C36}$ tilted 17° away from the lamella plane normal (corresponding SAED is shown in Figure 10 (c)), as well as the line direction of the intersections of the C36-dislocation slip habit planes with the lamella surface (indicated in Figure 10 (a) with blue dashed lines).

Further insights into the crystallographic orientation of the related slip system were obtained with analysis of a second FIB lamella cut from the sample volume indicated in Figure 9 with a white dotted rectangle, i.e. within the material volume located beneath the indent and parallel to the indentation direction but perpendicular to both the sample surface and the deformation sub-surface. The composite microstructure present in the lamella is shown in Figure 11 (a). Again, matrix regions located on either side of the Laves phase belong to the same Mg grain, exhibiting accordingly the same crystallographic orientation. The C36 Laves phase consisted also of a single grain across the lamella. The local orientation relationship between the Mg matrix and the Laves phase identified was very similar as found for the lamella cut from the surface (Figure 10). Basal planes of both hexagonal phases exhibited again almost perpendicular relative orientation, with $\angle[0\,0\,0\,1]_{C36}, [1\,0\,\bar{1}\,0]_{\alpha\text{-Mg}} = 3°$ and $[1\,\bar{2}\,1\,0]_{C36} \parallel [\bar{1}\,2\,\bar{1}\,0]_{\alpha\text{-Mg}}$. A similar orientation relationship, specifically $\angle[0\,0\,0\,1]_{C36}, [0\,0\,0\,1]_{\alpha\text{-Mg}} = 82° \pm 10°$, was also observed in several other FIB lamellae from alloys having slightly different compositions but similar microstructures, as shown in [64] and in Fig. SM 7. The presence of basal slip traces was identified in the Mg matrix as well as in the C36 phase. Figure 11 (b) and

(c) show magnified BF images of the Mg matrix and C36 phase, at the extreme sample tilt conditions, corresponding to B ∥ $[1\,1\,\bar{2}\,0]_{\alpha\text{-Mg}}$ (in (b)) and B ∥ $[1\,0\,\bar{1}\,0]_{C36}$ (in (c)). Basal-plane stacking faults of both phases are observed in the respective images in edge-on orientations (indicated in (b) and (c) with yellow arrows). Corresponding SAED patterns are presented in (d) and (e). It is important to note, that BF images of planar defects in their edge-on orientations were acquired at extreme α and β tilt values of the TEM sample holder - more than 80° apart from each other (at the very limits of the TEM holder tilt capabilities). Accordingly, the projection thickness of the FIB lamella was exceptionally high, resulting in a relatively poor quality of the BF TEM images. Nonetheless, the presence of dislocations evidencing the activation of a non-basal deformation system within the α-Mg matrix was also observed here. Sets of dislocations piled-up at the α-Mg/Laves phase interface, confined within non-basal slip planes of the α-Mg grain are indicated in Figure 11 (b) with blue arrows, while the line directions of the intersection of their slip planes with the lamella surface are approximately indicated by blue dashed lines.

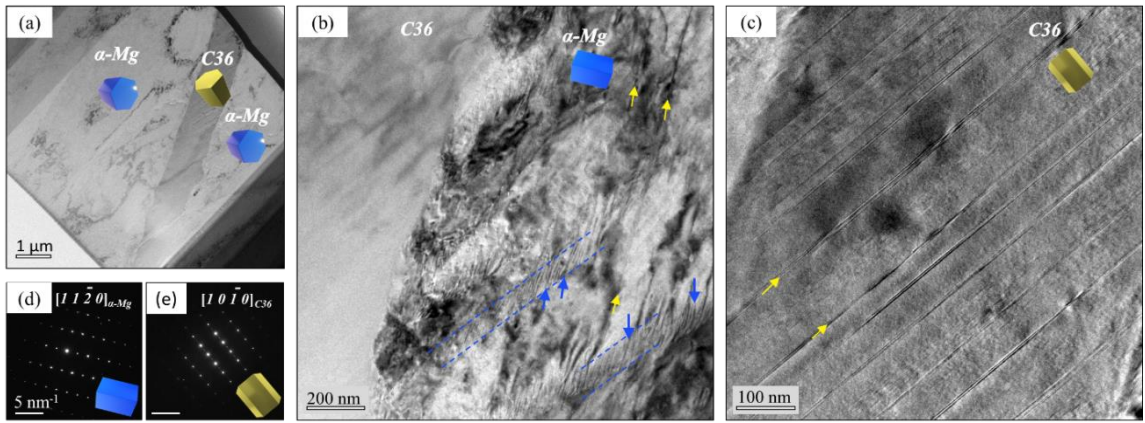

*Figure 11: Composite microstructure of the FIB lamella, observed in TEM: (a) BF overview image in the initial lamellae orientation (without any sample tilt). b) Magnified BF appearance of deformed α-Mg matrix in the lamella orientation B ∥ $[1\,1\,\bar{2}\,0]_{α\text{-}Mg}$. Basal-plane SFs (edge-on) and non-basal dislocations of the Mg matrix are indicated in (b) with yellow and blue arrows, respectively. c) Basal-slip SFs in C36 Laves phase (yellow arrows), observed in edge-on orientation at B ∥ $[1\,0\,\bar{1}\,0]_{C36}$. d)-e) SAED patterns corresponding to the Mg matrix at B ∥ $[1\,1\,\bar{2}\,0]_{α\text{-}Mg}$ and Laves phase at B ∥ $[1\,0\,\bar{1}\,0]_{C36}$, as presented in (b) and (c), respectively.*

## 4. Discussion

### 4.1. Nanoindentation studies

#### 4.1.1. CSR tests

The RT hardness measured across α-Mg-Laves phase interface regions is higher than that of the α-Mg phase (which amounts to ≈ 0.95 GPa, see Figure 2). In the case of indents specifically at the interface, the higher measured hardness is largely attributed to the higher hardness of the Laves phase when compared to the soft α-Mg phase, i.e. it occurs from a rule-of-mixtures effect, rather than strengthening resulting from the interfaces inhibiting dislocation motion. However, due to the complex microstructure and three-dimensional stress field from indentation, it is not possible to determine the specific volume fractions of each phase deformed by each indent and examine this phenomenon in more detail.

Specifically, the hardness of the $Mg_2Ca$ phase was measured to be ≈ 3.42 GPa at RT [28], nearly 3.5 times higher than that of the α-Mg phase (≈ 0.95 GPa at RT). No drop in hardness of the $Mg_2Ca$ phase was observed until 200 °C (see Figure 2 (a)). This is somewhat consistent with the results of Kirsten et al. [65], who also reported only slight loss in the Brinell hardness with temperature for the $Mg_2Ca$ phase up to the ductile-to-brittle-transition temperature (0.59 $T_m$, where $T_m$ is the melting temperature). The hardness measured across the α-Mg/Laves phase interfaces has the similar level at RT and at 170 °C. This clearly indicates that the Laves phase retains its hardness at least until 170 °C during the short interval CSR testing.

However, as expected, the hardness of the α-Mg phase decreases with temperature (see Table 1). It is therefore assumed that the superior creep properties of Mg-Al-Ca alloys when compared to conventional Mg-Al-Mn or AZ91 alloys [5-7] is caused by the fact that the Laves phases maintain their hardness with temperature up to at least 200 °C.

Moreover, it is clear from Figure 2 (a and b) that the hardness of the α-Mg matrix decreases with temperature and also that an apparent indentation size effect (at depths below 1000 nm) is observed for the α-Mg matrix at all test temperatures. The indentation size effect is attributed to the density of GNDs generated due to the strain gradient underneath the indenter [44]. The effect is more pronounced at low depths, which are usually encountered in nanoindentation [44, 46]. However, since the indentation size effect is not the focus of this study, it is not further discussed here. It can also be seen in Figure 2 (b-d), that there is a slight but noticeable increase in mean hardness for the α-Mg phase after an indentation depth of ~1000 nm was reached. This is assumed to be caused by the back stress work hardening associated with the pile-up of dislocations at α-Mg/Laves phase interfaces surrounding the indent location at a greater distance.

### 4.1.2. SRJ and creep tests

The strain rate sensitivity and activation volume are two important parameters that are connected to the deformation mechanisms of metallic materials [66]. Materials with a high strain rate sensitivity have a smaller activation volume and hence a higher probability of thermal activation of plasticity mechanisms [51, 67]. The activation volume of fine-grained Mg alloys at room temperature was calculated to be within the range of 20-80 $b^3$ by Somekawa et al. [67], who concluded that the rate controlling mechanism was most likely dislocation cross-slip. Somekawa et al. [54] also carried out nanoindentation studies on coarse grained (~100 μm) pure Mg and Mg-0.3 at.% (Al, Li, Y and Zn) alloys at room temperature. They calculated activation volumes in the range of 45-105 $b^3$ depending on the different compositions. In these coarse grained Mg alloys, deformation twins were visible in the regions around indents, and it was assumed that these deformation twins play a similar role in coarse grained materials as grain boundaries in fine grained materials: acting as obstacles for dislocations [54]. Table 2 summarizes activation volumes and the associated mechanisms of different Mg alloys reported in the literature.

*Table 2: Activation volumes and associated deformation mechanisms of pure Mg and Mg alloys.*

| Testing method | Material | Grain size | Activation volume ($b^3$) | Proposed mechanisms |
|---|---|---|---|---|
| Compression [68] | Pure Mg | 120 μm | 409 | Forest dislocation activity |
| Compression [68] | Pure Mg | 400 nm | 42 | Twinning |
| Compression [68] | Pure Mg | 60 nm | 12 | Grain boundary sliding |
| Nanoindentation [67] | Pure Mg | ~2-3 μm | 25.3 | Cross-slip |
| Nanoindentation [67] | Mg-0.31at% Al | ~2-3 μm | 80.8 | Cross-slip |
| Nanoindentation [67] | Mg- 0.3at% Ca | ~2-3 μm | 80.8 | Cross-slip |
| Nanoindentation [54] | Pure Mg | 101.9 μm | 45 | Cross-slip |
| Nanoindentation [54] | Mg-0.3 at.% Al | 83 μm | 105 | Cross-slip |
| Nanoindentation [54] | Mg-0.3at.%Zn | 90.3 μm | 93 | Cross-slip |
| Nanoindentation SRJ [53] | ZK 60 (Mg-5.5Zn-0.5Zr) | ~20 μm (25%) ~1-1.5μm (75%)* | 7.38 | Grain boundary sliding |
| Tensile [69] | Pure Mg | 2.1 μm | 9.8 at <$10^{-4}$/s & 15.2 at >$10^{-4}$/s | Grain boundary sliding at <$10^{-4}$/s and cross-slip at >$10^{-4}$/s |
| Tensile [69] | Pure Mg | 18.5 μm | 20.1 at <$10^{-4}$/s & 23.9 at >$10^{-4}$/s | Cross-slip and multiple-slip |

*extruded alloy which was subjected to high pressure torsion resulting in a bi-modal grain size distribution.

In the present study, the strain rate sensitivity of the α-Mg phase marginally increased from 0.013 ± 0.06 to 0.023 ± 0.009 and the activation volume slightly decreased from 86 ± 34 $b^3$ to 70 ± 26 $b^3$, when the temperature was increased from 100 °C to 170 °C. The values calculated in the present work are close to the values reported by Somekawa et al. [54, 67] for fine and coarse grained Mg alloys. As mentioned, the plastic deformation of Mg is facilitated by deformation twinning and dislocation slip. Deformation twinning is not a thermally activated process but is important for the deformation of Mg alloys [54, 69, 70]. However, the strain rate sensitivity and activation volume values obtained from nanoindentation reveal thermal activation of the α-Mg phase. This means that in addition to the observed deformation twinning, thermally-activated deformation mechanisms are also active in the α-Mg phase. The $V^*$ values obtained from nanoindentation SRJ tests are below 100 $b^3$ (at temperature ≥ 100 ºC) indicating thermally activated dislocation cross-slip in the α-Mg phase similar to what has been reported by Somekawa et al. [54, 67].

The activation volume of α-Mg/Laves phase ((Mg,Al)$_2$Ca and Mg$_2$Ca) interfaces was slightly lower than that of the α-Mg phase at all testing temperatures (100, 140 and 170 ºC, see Table 1). The hardness of the Mg$_2$Ca phase stays constant at least till 200 ºC, which indicates negligible thermal activation in the temperature range investigated for this phase. However, a lower activation volume of α-Mg-Laves interfaces indicates that thermally activated mechanisms like interfacial sliding may be active to some extent in these interfacial regions. The interfacial sliding in this alloy system has been reported earlier [30, 31, 71].

It is evident from Fig. SM 5 that the N value for the Mg$_2$Ca phase appears to change at low strain rates (~ ≤ 0.01/s, i.e. in the creep experiments) and a temperature of 170 °C. The N value for the Mg$_2$Ca phase (at low strain rates) is nearly the same as that of the α-Mg phase. Further, a loss in hardness with holding time is also observed at low strain rate for the $Mg_2Ca$ phase. This is consistent with the results published by Rokhlin et al. [29], who reported a loss in microhardness with increasing holding time at temperatures ≥ 150 °C. Unfortunately, our data for the $Mg_2Ca$ phase (as can be seen in the Fig. SM 5) is affected by thermal drift. Therefore, we did not attempt to derive thermally activated mechanisms active in the $Mg_2Ca$ phase based on these experiments.

Similar to what was observed during microindentation (Figure 6 and Figure 9), parallel slip lines on the surface of Laves phases (particularly in $(Mg,Al)_2Ca$) were also observed in the regions around nanoindents, indicating that deformation is confined to particular and equivalent crystallographic planes.

**4.2. Co-deformation mechanisms**

The CRSS values of all available slip systems differ highly for the α-Mg matrix and the Laves phases (see Fig. SM 1). It therefore might be expected that the probability of slip transfer from one phase to another is low. However, slip transfer from the soft α-Mg phase to Laves phase is exactly what is observed in this work.

Generally, this unexpected slip transfer can be attributed to two main reasons. Firstly, the slip behaviour observed from individual single crystal experiments cannot truly account for polycrystalline alloys. Specifically, in the case of polycrystalline Mg alloys, the presence of grain boundaries [72, 73], alloying elements [74] and precipitates [75, 76] dramatically affects the deformation behaviour and activation of non-basal slip systems. Moreover, the CRSS ratio between non-basal and basal slip systems estimated in polycrystalline Mg alloys is quite low (~2-3) in comparison to the estimates deduced from single crystal testing (~80-100) (Fig. SM 1). Therefore, non-basal slip is also readily observed at room temperature in several Mg alloys [77-79].

Secondly, stress concentrations caused by the pile-up of dislocations at the phase boundaries between the soft matrix and the hard Laves phases can promote slip transfer. Consequently, it is assumed that during indentation, basal dislocations in the α-Mg matrix start to glide first (see Figure 6 and Figure 7, where basal slip lines are readily evident around indents). However, their movement is blocked at the (C36) Laves phase boundaries due to the different CRSS, crystal structure, nature of dislocations and crystal orientation (see Figure 10 and Figure 11) of (C36) Laves phase and the α-Mg matrix. The blocking of dislocations creates pile-ups of dislocations at the phase boundaries and an associated strain gradient, shown by the increasing hardness of deep indents (Figure 2 (b)) [80, 81]. Phase boundaries are generally considered to be stronger obstacles to dislocation movement than grain boundaries [82-84]. This is because the transmission of dislocations across phase boundaries requires in most cases the nucleation of new dislocations due to the different crystal structures, crystal orientations and lattice parameters of both phases [82-84]. The pile up of dislocations will then create back stress work hardening in the α-Mg matrix. The compatibility and hardening effect by the dislocations pile ups can then activate several favourably orientated non-basal dislocations sources in the α-Mg matrix and at the phase boundaries. The pile up of non-basal dislocations at the α-Mg/Laves phase boundary is marked by blue arrows in Figure 11 (b).

In addition to back stress work hardening of the soft phase, the pile up of basal and non-basal dislocations at the interface will create forward stresses in the C36 Laves phase [80, 81]. The magnitude of the forward stress in the intermetallic phase can be several times higher than the applied stress and is usually calculated as

$$n \cdot \tau_a \qquad \qquad Eq.\ 5$$

where $\tau_a$ is the applied shear stress and *n* is the number of dislocations in the pile-up [81, 85]. A high stress concentration can then trigger the activation of basal dislocations in the C36

Laves phase to release the stress concentration at the phase boundaries. A recent study by Guénolé et al. [64] presents the slip transfer mechanisms from the α-Mg phase to the Mg$_2$Ca Laves phase in a situation where both phases are aligned perpendicular to each other. They [64] found that under the application of a compressive stress, plasticity is nucleated in the α-Mg matrix and then the local stress concentrations at the α-Mg/Laves phase interface can induce the nucleation of prismatic <a> dislocation from the interface. The present work reports a similar orientation relationship between the α-Mg phase and the C36 Laves phase (see Figure 10 and Figure 11). However, in this work, basal slip, rather than prismatic slip, was observed in the C36 Laves phase.

The role of non-basal dislocations seems to be important for the activation of basal dislocation slip in the C36 Laves phase because of geometrical reasons. One is the very simple geometric compatibility factor widely used to estimate the ease of slip transfer. This Luster-Morris parameter, $m'$, suggested by Luster and Morris, is calculated as (Eq. 6) [86]:

$$m' = \cos\Phi . \cos\kappa \qquad \text{Eq. 6}$$

where $\Phi$ and $\kappa$ are the angle between the slip plane normals and the slip directions in two neighbouring grains or phases, respectively. An $m'$ value close to 1 indicates that the slip systems are crystallographically aligned and highly favourable for slip transmission, while for $m' = 0$ the slip systems are completely incompatible. When applying this criterion to α-Mg/Laves phase interfaces, it is highly unlikely that basal slip in the α-Mg matrix will trigger basal slip in the C36 phase. This is because the basal plane normals of both phases, $\Phi$, are at an angle of ~90 °, causing a low $m'$ value. However, the pile-up of non-basal dislocations can create relatively favourable stress concentrations at the α-Mg/Laves phase interface, which can trigger the activation of basal slip sources in the C36 phase. The proposed mechanism is shown schematically in Figure 12. Zhu et al. [76] also reported the presence of non-basal geometrically necessary dislocations (GNDs) in the Mg phase around $Al_2Ca$ Laves phase precipitates. They also reported slip on {111} planes in the $Al_2Ca$ phase.

In addition to slip transfer, crack formation in the Laves phase is another way of releasing stress concentrations at the phase boundary. The observed different plastic behaviour such as dislocation slip or crack formation in the Laves phase is assumed to occur because of the following three reasons when considering the observed orientation relationship:

1. The orientation of the cracks (see Figure 9 d) in the Laves phase is often parallel to the basal plane indicating that the crack might have formed after some plastic deformation on the basal plane [64]. For the crystallographically related µ-phase, similar decohesion along planes with high dislocation density after plastic deformation was observed in deformed micropillars by TEM [87].
2. Due to the preferred orientation relationship between the α-Mg matrix and the C36 Laves phase, the activation of non-basal slip in the α-Mg phase seems critical for initiating slip in the Laves phase. Therefore, α-Mg grains having a high Schmid (or Taylor) factor for non-basal slip systems can facilitate slip transmission from Mg into the Laves phases. Crack formation in the Laves phase would then be more pronounced, when non-basal slip in the α-Mg phase cannot be activated.
3. The only mode of plastic deformation of the Laves phase was basal, therefore, the Schmid factor for basal slip in the C36 Laves phase is also of significant importance. A C36 Laves phase loaded having a low Schmid factor for basal slip will have low chances to undergo plastic deformation and will be prone to crack formation.

Due to the complex stress state associated with indentation, however, the Schmid factors for several slip systems in this work could not be computed. The critical resolved shear stresses for activating slip in the C36 Laves through co-deformation could be obtained using micro-pillar compression in a future study.

Unfortunately, synthesis of a corresponding sample with large enough, single phase C36 grains for nanomechanical testing proved unsuccessful during this work and we therefore compare our results with the better known and structurally very similar hexagonal C14 phase. With suitable bulk C36 Laves phase material, nanomechanical tests could also be performed over a range of temperature to analyse the effect of temperature on its hardness and deformation behavior and to potentially separate this from any effects of changing composition and stacking fault density in the presence of a matrix phase. A similar challenge exists in simulations, where the availability of a ternary Mg-Al-Ca interatomic potential limits the use of MD simulations that could shed further light on dislocation mechanisms in the C36 phase and at the Mg-C36 interface. Until now, such computational work has been limited to the Mg-C14 system and therefore both, experiments and simulations, will benefit from progress in synthesis and empirical potentials to further reveal and separate the mechanisms we observe here.

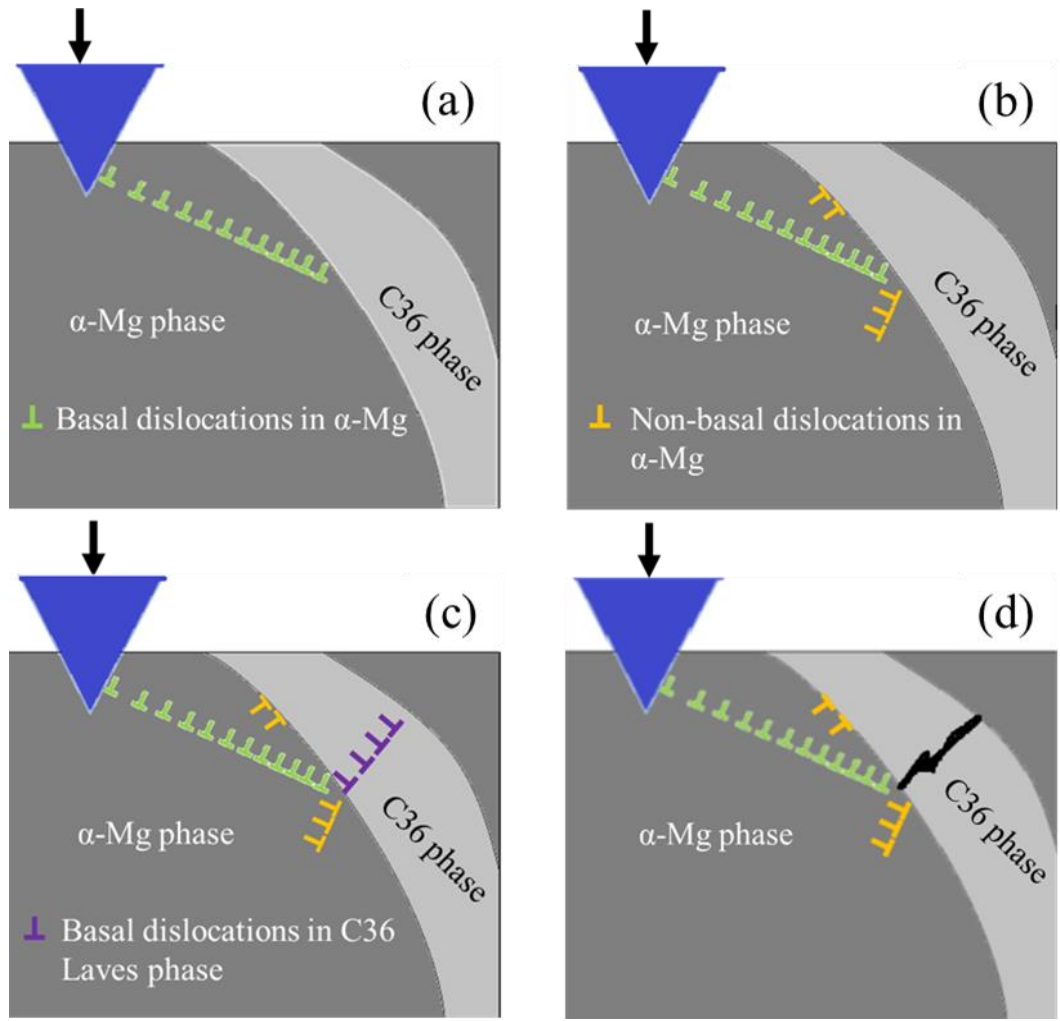

*Figure 12: Co-deformation mechanisms of soft α-Mg matrix with hard Laves phase, (a) basal slip in the α-Mg matrix and pile-up of dislocations at an α-Mg/Laves phase interface, (b) activation and generation of non-basal dislocations in the matrix close to the phase boundary, (c) deformation by basal slip in Laves phase or (d) cracking of the Laves phase at stress concentration points. (d) might occur after (c).*

### 4.3. Implications of the present work for alloy design

As described in the previous sections, the Laves phase is much harder and requires significantly higher stresses than the α-Mg matrix for plastic deformation. The presence of the Laves phase, therefore, significantly restricts the deformation occurring in the α-Mg matrix as is evident around the microindents (Figure 8). Further, it is clear from this study that the $Mg_2Ca$ phase does not lose hardness with temperature. This means the Laves phase can provide strengthening to the α-Mg matrix at elevated temperatures that are usually encountered during creep loading of automotive components, like powertrains [4, 88], and is most probably the reason for the good creep properties of these alloys.

There are also a few interesting prospects owing to the nature of co-deformation of the Laves phases in the metallic matrix, namely that they are apparently non-shearable obstacles to dislocations but at the same time can relieve the stress of dislocation pile-ups by nucleation and motion of dislocations inside the intermetallic phase itself. Zhu et al. [76] have recently used the non-shearable yet deformable nature of $Al_2Ca$ precipitates to design a Mg-6Al-1Ca alloy with significant ductility. They attributed the high work hardening capacity of the alloy to the increased density of geometrically necessary dislocations nucleated and accumulated around $Al_2Ca$ precipitates. Similarly, we also observed plastic deformation in the $(Mg,Al)_2Ca$ phase itself and non-basal dislocations around interfaces in this work.

As with the case of lean Ca alloys [76], in alloys with relatively high Ca content the deformable nature of the $(Mg,Al)_2Ca$ phase can then also be used as a tool to design creep resistant alloys that also exhibit sufficient ductility and work hardening capacity. Since the small size of the microstructure enables plasticity in the Laves phase [28, 89], there is additional scope for alloy design in terms of its microstructural length scales. Different alloys with varying thicknesses of intermetallic Laves phase struts (by alloying additions or cooling rate adjustments) can be produced and then the Laves phase thickness effects on slip transfer and mechanical properties can be studied and ultimately used to fine tune the alloy composition. These thoughts are summarised in the Figure 13.

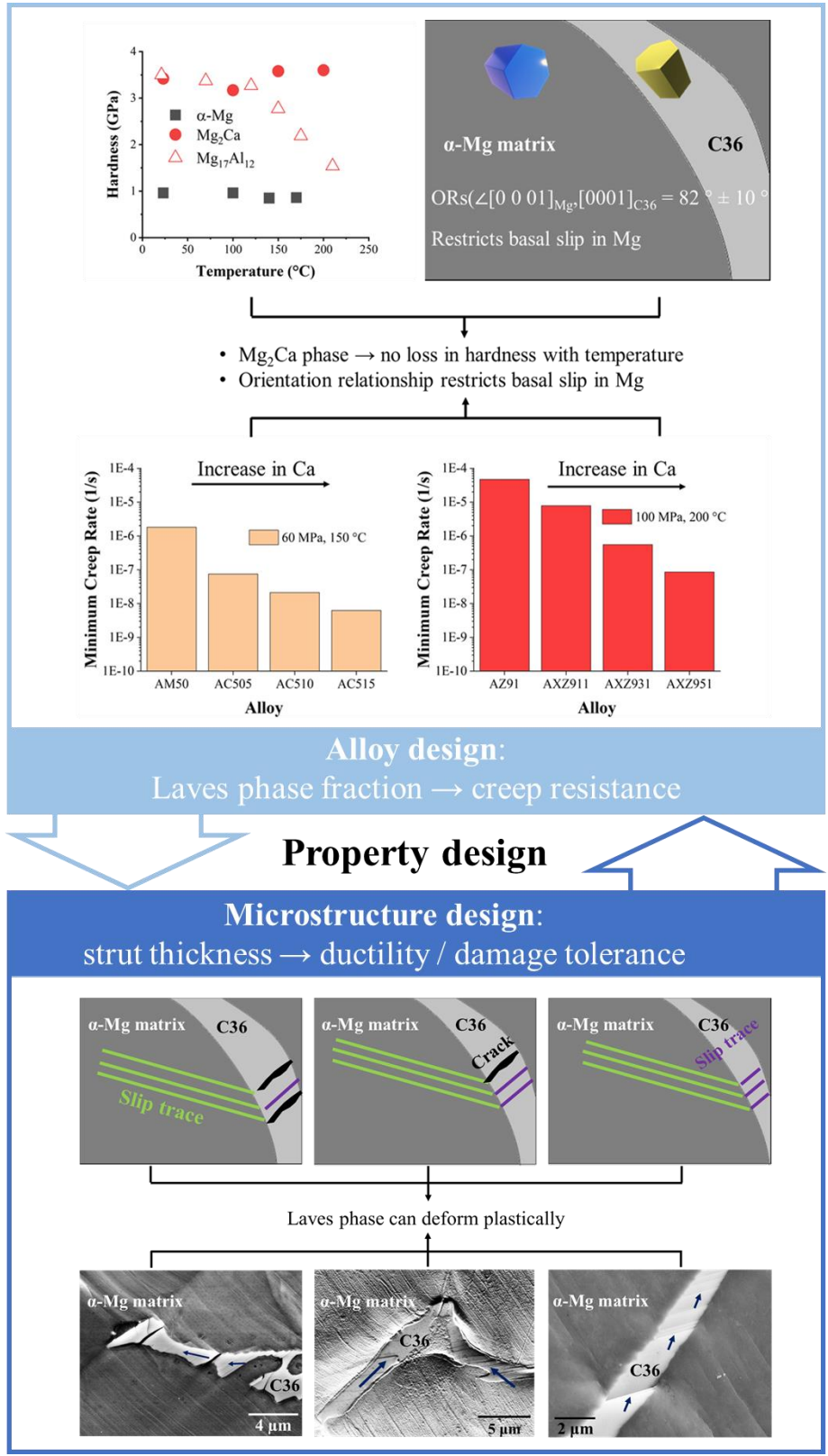

*Figure 13: Significance of the present work and possible future work directions. Average hardness of $Mg_{17}Al_{12}$ at different temperatures were drawn out from [57]. Values of minimum creep rates and tensile properties for AM50, AC505, AC510 and AC 515 alloys were extracted from [10], while for AZ91, AXZ911, AXZ931 and AXZ 951 alloys from [6].*

## 5. Conclusions

The goal of this work was to study the co-deformation behaviour of hard Laves phases in a soft magnesium matrix. In particular, it was necessary to determine the mechanical properties of the individual phases and thermally activated deformation mechanisms, elucidate the effects of orientation and Laves phase on the deformation of the α-Mg matrix, and study slip transfer from the soft α-Mg phase to the hard intermetallic Laves phase. Based on the results of nano- and microindentation experiments on alloys AX44 and Mg-30Ca, along with post-deformation SEM and TEM studies, the following conclusions can be drawn:

i)  The hardness of the α-Mg phase decreases slightly from 0.96 GPa to 0.86 GPa when the temperature increases from RT to 170 °C, while the hardness of the $Mg_2Ca$ phase stays approximately constant. The hardness across α-Mg-Laves phase interfaces was higher than that observed for the α-Mg phase at all testing temperatures.

ii) The strain rate sensitivity of the α-Mg phase slightly increases from $0.013 \pm 0.006$ at 100 °C to $0.023 \pm 0.009$ at 170 °C while its activation volume decreases from 2.84 $nm^3$ ($86 \pm 34$ $b^3$) to 2.31 $nm^3$ ($70 \pm 26$ $b^3$). These activation volumes indicate that deformation is dominated by dislocation cross-slip.

iii) The activation volume at α-Mg/Laves phase interfaces is smaller than that of the α-Mg phase suggesting that thermally activated mechanisms, like interfacial sliding, are active in the interfacial regions.

iv) The creep deformation of the $Mg_2Ca$ Laves phase at the same load and a temperature of 170 °C is significantly lower than that of the α-Mg phase.

v)  The deformation zone around and below indents in the α-Mg matrix is both orientation-dependent and strongly influenced by adjacent Laves phases. Sub-surface deformation mechanisms are well-represented by deformation around indents at the surface.

vi) Co-deformation of the Laves phase and the α-Mg matrix occurs, despite the Laves phase being significantly harder than the α-Mg phase. This is evidenced by slip lines in the Laves phase being observed in regions where slip lines or twins in the α-Mg matrix intersect with the Laves phase.

vii) A preferential orientation relationship, with basal (0001) planes of the constituent phases being almost perpendicular to each other, was revealed in the AX44 dual-phase material investigated by TEM.

viii) Deformation of the α-Mg phase occurs via mechanical twinning, basal and non-basal slip. However, basal slip was observed to be predominant. The C36 Laves phase deforms mainly by basal slip as was identified by TEM analysis.

**Acknowledgments**

The authors are immensely grateful for the financial support provided by the Deutsche Forschungsgemeinschaft (DFG), under Collaborative Research Centre (CRC) 1394, project ID 409476157 within projects A03, A05, C01 and C02. We are also thankful to Dr. Risheng Pei for his help and support in analysing EBSD data. Further, we also gratefully acknowledge the help received from our colleagues, Mr. David Beckers, Mr. Maximilian Kruth, Mr. Gerhard Schutz and Mr. Arndt Ziemons (in alphabetical order), at various steps in this study.

**Data availability**

The raw/processed data required to reproduce these findings cannot be shared at this time as the data also forms part of an ongoing study.